\documentstyle[preprint,eqsecnum,aps]{revtex}

\input psfig.tex

\newcommand{\be}{\begin{equation}}    
\newcommand{\ee}{\end{equation}}
\newcommand{\beq}{\begin{eqnarray}}
\newcommand{\eeq}{\end{eqnarray}}

\def\op{ \ $ }
\def\cl{$ \ }
\def\nn{\nonumber}

\def\ver{\vskip 12pt}

\def\NR{N_{\ell m}(\omega,r)}
\def\LR{L_{\ell m}(\omega,r)}
\def\VR{V_{\ell m}(\omega,r)}
\def\VRP{V'_{\ell m}(\omega,r)}
\def\VRPP{V''_{\ell m}(\omega,r)}
\def\hhr{h^{0}_{\ell m}(\omega,r)}
\def\hhrp{h^{0'}_{\ell m}(\omega,r)}
\def\h1r{h^{1}_{\ell m}(\omega,r)}
\def\ha1rp{h^{1'}_{\ell m}(\omega,r)}

\def\Y{Y_{\ell m}(\vartheta,\varphi)}

\def\Yt{\frac{\partial Y_{\ell m}(\vartheta,\varphi)}{\partial\vartheta }}
\def\Ytt{\frac{\partial^2 Y_{\ell m}(\vartheta,\varphi)}{\partial\vartheta^2 }}
\def\Yf{\frac{\partial Y_{\ell m}(\vartheta,\varphi)}{\partial\varphi }}
\def\EM{e^{-i\omega t}}

\def\N{N_{\ell m}(r,\omega)}
\def\T{T_{\ell m}(r,\omega)}
\def\V{V_{\ell m}(r,\omega)}
\def\L{L_{\ell m}(r,\omega)}
\def\hh{h^{0}_{\ell m}(r,\omega)}
\def\hs{h^{1}_{\ell m}(r,\omega)}

\def\ll{\left[}
\def\rr{\right]}

\def\gbl{\Biggl\{}
\def\gbr{\Biggr\}}

\def\ps{\Psi_{\ell m}^{BPT}(\omega,r)}
\def\pps{\Psi^{BPT}}
\def\zns{Z^{NS}_{\ell m}(\omega,r)}

\def\rwe{ Regge-Wheeler equation }
\def\bpte{ BPT equation }

\begin{document}


\draft

\title{Stellar pulsations excited by a scattered mass}

\author
{V. Ferrari$^{1}$, L. Gualtieri$^2$, A. Borrelli$^3$}

\address
{$^1$ Dipartimento di Fisica ``G.Marconi",
 Universit\` a di Roma ``La Sapienza"\\
and \\
Sezione INFN  ROMA1, p.le A.  Moro
5, I-00185 Roma, Italy\\
$^2$ Dip. di Fisica Teorica,
Universit\` a di Torino, 10125, Italy\\
$^3$ Dip. di Fisica,
Universit\` a di Pavia, Pavia, 27100, Italy}

\date{\today}

\maketitle

\begin{abstract}

We compute the energy spectra of the gravitational 
signals emitted when a
mass \op m_0\cl is scattered by the gravitational field of a star of mass
\op M >> m_0.\cl We show that, unlike black holes in similar processes,
the quasi-normal modes of the star are 
excited,  and that the amount of energy emitted in these modes depends
on how close the exciting mass can get to the star.

\end{abstract}
 
\pacs{PACS numbers: 04.30.-w,  04.40.Dg}

\narrowtext

\section{Introduction}

\ver\ver
In this paper we study the gravitational signals
emitted when a mass \op m_0\cl is scattered by the gravitational field of
a star of mass \op M.\cl 
Under the assumption that \op m_0\cl is much smaller than \op M,\cl this 
study can be done in the framework of a first order perturbation theory,
by assuming that the perturbations of the gravitational
field and of the fluid composing the star are excited by the 
stress-energy tensor of a
point-like scattered mass  (a {\it particle}).

Since 1971, when M.Davis, R.Ruffini, W.H.Press and R.H.Price  
\cite{davisruffinipressprice}  computed the energy 
and the waveform emitted when a particle
is radially captured by a Schwarzschild black hole,
the perturbations of  rotating and nonrotating black holes
excited by infalling,  scattered, or
orbiting masses, have been extensively investigated. 
No comparable attention has been dedicated to  the
study of stellar perturbations  excited by small masses,
although this problem is certainly interesting.
Indeed, we know that neutron stars exist, and observations allow
to infer the location and the mass  of many of them, as well as their
rotation periods. In addition, the worldwide experimental effort 
spent to detect gravitational waves, will hopefully be crowned by
success  in a not too far future.
Therefore, it is time to gain new theoretical
insight on  the characteristics of the gravitational signals emitted by
neutron stars.
Until now, the application of the theory of stellar perturbations 
has mainly regarded the determination  of the
frequencies of the quasi-normal modes, at which stars  are expected to
pulsate and  emit gravitational waves. 
However,  whether these modes can be excited,
and how much energy  is emitted at the corresponding
frequencies, can be understood only by considering
realistic astrophysical situations.
The theory of stellar perturbations, suitably matched with the theory 
of black hole perturbations, allows   to  explore this new field.

In our investigation we shall consider a polytropic star 
with mass and radius typical of  neutron stars,
and integrate the equations describing the scattering of a small mass.
In the interior of the star, we shall use the equations 
for the axial and polar  perturbations derived by 
S.Chandrasekhar \& V.Ferrari
in 1990 in ref. \cite{chandraferrari1},
(to be referred to hereafter  as Paper I), that hold also
in a more general gauge appropriate to describe
non-axisymmetric perturbations.
Outside the star, one may  continue the solution by  integrating the
Regge-Wheeler and the Zerilli equations \cite{reggewheeler},
\cite{zerilli},  with a source term given by the
stress energy tensor of the scattered mass moving  along a geodesic 
of the unperturbed spacetime. However,  it is known
that the source term of these equations
diverges when the mass reaches the periastron, 
and therefore we  switch to a different formalism, and   
use the generalized non-homogeneous 
Regge-Wheeler equation, which was introduced by
T.Nakamura and M.Sasaki \cite{nakamurasasaki}
to overcome this problem. We shall  integrate  this
equation by adopting the procedure discussed in ref.
\cite{ferrarigualtieri} (to be referred to hereafter  as Paper II),
and further developed in this paper.  
We shall show how the  problem of matching the interior and exterior
solution can be solved, and find the energy spectra  of the emitted
radiation for different values of the orbital parameters of the scattered
mass. 

The plan of the paper is the following. In section II we shall briefly
review the equations to be integrated in the interior of the
star;
in section III the equations governing the
exterior perturbations will be shown; 
the procedure to find the complete solution and the matching 
conditions will be discussed in section IV;
in section V we will describe a  model of polytropic star 
for which we shall compute the characteristic oscillation frequencies;
in section VI we will show the energy spectra emitted 
when a small mass is scattered by such star.

\section{The equations describing the perturbed spacetime inside the star.}

In order to describe the perturbations induced by a mass scattered by a
spherical, nonrotating  star, we  choose a gauge appropriate to describe 
non-axisymmetric perturbations
\beq
\label{pertmetric}
ds^2 &=&
e^{2\nu(r)} dt^2 - e^{2\mu_2(r)} dr^2 -
r^2 d\vartheta^2-r^2\sin^2\vartheta d\varphi^2\\
\nn
&+& \sum_{\ell m}~~\int^{+\infty}_{-\infty} d\omega~\EM
\Biggl\{
2 e^{2\nu}\N\Y dt^2-
2 e^{2\mu_2}\L\Y dr^2\Biggr.\\
\nn
&-&\Biggl. 2 r^2 H_{33} d\vartheta^2
- 2 r^2\sin^2\vartheta 
H_{11} d\varphi^2
- 2r^2H_{13} d\vartheta d\varphi 
+ \hh\sin\theta\Y dt d\varphi\Biggr. \\
\nn
& +&\Biggl. \hs\sin\theta\Yt dr d\varphi
- \hh\frac{1}{\sin\theta}\Yf dt d\vartheta\Biggr. \\
\nn
&-&\Biggl. \hs\frac{1}{\sin\theta}\Yf dr d\vartheta
\Biggr\}
\eeq
\\
where  
\beq
\label{hhh}
\nn
   H_{11}&=&\left[T_{\ell m}(r,\omega)+
V_{\ell m}(r,\omega) \left( \frac{1}{\sin^2{\vartheta}}
     \frac{\partial^{2}}{\partial \varphi^{2}}
     +\cot{\vartheta}\frac{\partial}{\partial \vartheta}
     \right) \right] \ Y_{\ell m}(\varphi,\vartheta) \\\nn
   H_{13}&=&V_{\ell m}(r,\omega)\left[
\frac{\partial^2}{\partial\varphi\partial\vartheta}
-\frac{\partial}{\partial\varphi}\cot\vartheta\right] Y_{\ell m}\\\nn
   H_{33}&=&\left[T_{\ell m}(r,\omega)+
V_{\ell m}(r,\omega)\frac{\partial ^{2}}
     {\partial \vartheta^{2}}\right]\ Y_{\ell m}(\varphi,\vartheta) \ .
\nn
\eeq
and \op \Y\cl are the scalar spherical harmonics.
The perturbed part of the metric (\ref{pertmetric})
has been Fourier-expanded,
and the usual decomposition in
tensor spherical harmonics has been performed 
(cfr. \cite{reggewheeler,zerilli}). 
The functions \op \Bigl[\N, \L, \V,\T\Bigr], \cl are the radial part
of  the {\it polar}, ({\it even}) metric components,
whereas \op \hh\cl and \op\hs\cl are  the {\it axial} ({\it odd})
part.
It should be mentioned that in the axisymmetric case ($m=0$) the previous
gauge reduces to that used in Paper I for the polar 
perturbations, and to the Regge-Wheeler gauge
\cite{reggewheeler} for the axial ones.

The unperturbed metric
functions \op\nu(r)\cl and \op \mu_2(r)\cl have to be determined by
numerically integrating the Einstein equations
coupled to the  equations of hydrostatic equilibrium,  for an
assigned equation of state.
Under the assumption that the star is composed by a perfect fluid
with energy-momentum tensor  given by
\op
T^{\alpha\beta}=(p+\epsilon)u^{\alpha}u^{\beta}-pg^{\alpha\beta},
\cl
where $p$ and $\epsilon$ are, respectively, the pressure and the energy
density, the relevant equations are
\beq
&&\left[ 1-\frac{2m(r)}{r}\right] p_{,r}  =  -(\varepsilon+p)\left[
 pr+\frac{m(r)}{r^{2}}\right],\\
&& \nu =-\int^r_0 {\frac{p_{,r}}{(\epsilon +p)}dr }+\nu_0,
\qquad\qquad
e^{2\mu_{2}}  =  \Bigl(1-\frac{2m(r)}{r}\Bigr)^{-1},
\label{einsteq}
\eeq
where \op m(r)=\int^{r}_{0}\varepsilon r^{2}dr\cl is the mass contained
within a sphere of radius $r$.
The solution for the function \op\nu(r)\cl requires the determination
of the constant $\nu_0$, which can be found by imposing that at 
the boundary of the star
the metric reduces to the Schwarzschild  metric
 \be
 (e^{2\nu})_{r=R}= (e^{-2\mu_2})_{r=R}=1-2M/R,
 \label{boundary}
 \ee
 where $M=m(R)$ is the total mass.

It is easy to check that, 
since \op Y_{\ell m}(\vartheta,\varphi)=
     (-1)^{m}Y^\ast_{\ell -m}(\vartheta,\varphi),\cl
the perturbed metric functions satisfy the following property
\be
\label{prop2}
F_{\ell m}(r,\omega)=(-1)^{m}F^{*}_{\ell -m}(r,-\omega),
\ee
where \op F_{\ell m}(r,\omega)\cl is any of the functions 
\op \Bigl[\N, \L, \V,\T\Bigr],\cl and \op\Bigl[\hh,\hs\Bigr].\cl

By writing explicitely the perturbed Einstein's equations coupled to the
hydrodynamical equations in the interior of the star,
it is possible to verify that the resulting separated
equations coincide with those given in Paper I, both for the polar and for
the axial perturbations.   As a consequence, the decoupling of the 
gravitational perturbations from the perturbations of the thermodynamical
variables that was performed in  Paper I
for the polar equations, is possible also in the present 
case of non-axisymmetric perturbations.
Therefore, the equations to integrate inside the star, to find the 
values of the polar and  axial functions  at the boundary, are the following.  
If  we consider stars with a barotropic
equation of state,  the  polar equations are
(cfr. eqs. (72)-(75)  of Paper I)
\beq
\label{poleq}
&&X_{,r,r}+\left(\frac{2}{r}+\nu_{,r}-\mu_{2,r}\right)X_{,r}+
\frac{n}{r^{2}}
e^{2\mu_{2}}(N+L)+\omega^{2}e^{2(\mu_{2}-\nu)}X=0,
\\\nn
&&(r^{2}G){,r}=n\nu_{,r}(N-L)+\frac{n}{r}(e^{2\mu_{2}}-1)(N+L)+r(\nu_{,r}
-\mu_{2,r})X_{,r}+\omega^{2}e^{2(\mu_{2}-\nu)}rX\; ,
\\\nn
&&-\nu_{,r}N_{,r}=-G+\nu_{,r}[X_{,r}+\nu_{,r}(N-L)]+
+\frac{1}{r^{2}}(e^{2\mu_{2}}-1)(N-rX_{,r}-r^{2}G)\\
\nn
&&-e^{2\mu_{2}}(\epsilon+p)N
+\frac{1}{2}\omega^{2}e^{2(\mu_{2}-\nu)}
\left\{ N+L+\frac{r^{2}}{n}G+\frac{1}{n}[rX_{,r}+(2n+1)X]\right\},
\\\nn
&&-L_{,r}=(N+2X)_{,r}+\left(\frac{1}{r}-\nu_{,r}\right)(-N+3L+2X)+\\
\nn
&&+\left[\frac{2}{r}-(Q+1)\nu_{,r}\right]\left[ N-L+\frac{r^{2}}{n}G+
\frac{1}{n}(rX_{,r}+X)\right]\; ,\nonumber
\eeq
where \op Q=\epsilon_{,r}/p_{,r},\cl the function \op V\cl has
been replaced by $X=nV$, with \op n=(\ell -1)(\ell+2)/2$, and
the function \op G\cl takes the place of \op T\cl
\beq
\label{defg}
G=\nu_{,r}[\frac{n+1}{n}X-T]_{,r}&+&
\frac{1}{r^{2}}(e^{2\mu_{2}}-1)[n(N+T)+N]+\frac{\nu_{,r}}{r}(N+L)-
\nonumber\\
&-&e^{2\mu_{2}}(\epsilon+p)N+
\frac{1}{2}\omega^{2}e^{2(\mu_{2}-\nu)}[L-T+\frac{2n+1}{n}X]\: .
\eeq
Outside the star, where \op\epsilon\cl and \op p\cl vanish,
these equations  become
appropriate to describe the Schwarzschild perturbations, and can be 
reduced to the Zerilli equation \cite{zerilli}
\be
\label{zereq}
\frac{d^{2}Z^{pol}}{dr_{*}^{2}}+
\Biggl\{\omega^{2}-\frac{2(r-2M)} {r^{4}(nr+3M)^{2}}
[n^{2}(n+1)r^{3}+3Mn^{2}r^{2}+ 9M^{2}nr+9M^{3}]\Biggr\}
Z^{pol}=0.
\ee
The value  of the Zerilli function  at the surface,
\op Z^{pol}(\omega, R),\cl
can be found in terms of the solution of the  eqs. (\ref{poleq}) 
as follows (cfr. eq. (93) Paper I)
\be
\label{zerfun}
Z^{pol}(\omega, R)=\frac{R}{nR+3M}\Bigl(
\frac{3M}{n}X(\omega,R)-RL(\omega,R)
\Bigr),
\ee
and similarly  for its first derivative.
For simplicity, in eqs. (\ref{poleq}-\ref{zerfun})
we have omitted the harmonic indices \op(\ell m).\cl

The equations for the axial  perturbations can be reduced to
a single wave equation
(cfr. eqs.  (148)and (149), Paper I)
\be
\label{axeq}
\frac{d^2Z^{ax}_{\ell m}}{dr_*^2}+\Biggl\{\omega^2-
\frac{e^{2\nu}}{r^3}\Biggl[\ell(\ell+1)r+r^3(\epsilon-p)-6m(r)\Biggr]\Biggr\}
Z^{ax}_{\ell m}=0,
\ee
where \op r_*=\int_0^re^{-\nu+\mu_2}dr,\cl
and the function \op Z^{ax}_{\ell m}\cl is related to the axial metric
components by
\be
h^0_{\ell m}=
\frac{2i}{\omega}\frac{d}{dr_*}\left( r Z^{ax}_{\ell m}\right) ,
\qquad\qquad
h^1_{\ell m}=2 e^{-2\nu}\left( r Z^{ax}_{\ell m}\right) .
\ee
Outside the star eq. (\ref{axeq}) reduces to the Regge-Wheeler equation.

\section{The equations describing the perturbed spacetime outside the star.}
The metric exterior to a nonrotating star reduces to the
Schwarzschild metric, and therefore the perturbed spacetime is described
by the perturbations of  a Schwarzschild  black hole. 
When the source exciting the perturbations is a scattered mass, the
source term of the Regge-Wheeler and of the Zerilli equations 
diverge at the periastron (where \op \frac{dr}{d\tau}=0$), 
and alternative equations have to be used.
In 1973  J.M. Bardeen and W.H. Press derived a  wave equation with a  
complex potential barrier for the
Weyl scalars \op\delta\Psi_0\cl and \op r^4\delta\Psi_4\cl
\cite{bardeenpress}. This equation 
was subsequently generalized by S.A. Teukolsky \cite{teukolski},
who found a master equation governing the electromagnetic,
gravitational and neutrino-field perturbations of a
Kerr black hole, which reduces to the Bardeen-Press equation
when the rotation parameter is set to zero.
We shall refer to the Bardeen-Press-Teukolsky equation as
the BPT equation.  After separating the variables, the radial BPT equation
can be written as
\be
\label{teukolsky}
\left\{
\Delta^2\frac{d}{dr}\left[\frac{1}{\Delta}\frac{d}{dr}\right]+
\left[\frac{\left(r^4\omega^2+4i(r-M)r^2\omega\right)}{\Delta}
-8i\omega r-2n\right]\right\}\pps_{\ell m} (\omega, r)= -T_{\ell m}(\omega, r),
\ee
where \op \Delta=r^2-2Mr,\cl the BPT function is related
to \op \delta\Psi_4\cl by 
\be
\label{psiquattro}
\delta\Psi_4(t,r,\vartheta,\varphi)=
\frac{1}{r^4}~\sum_{\ell m}~\int^{+\infty}_{-\infty}
d\omega~e^{- i\omega t}~_{-2}S_{\ell m}(\vartheta,\varphi)
\pps_{\ell m} (\omega, r),
\ee
and $ _{-2}S_{\ell m}(\vartheta,\varphi),\cl in terms of which the 
separation  is accomplished, is the
spin-weighted spherical harmonic
\beq
_{-2}S_{\ell m}(\vartheta,\varphi)=
\frac{1}{2\sqrt{n(n+1)}}\Biggl[
\Ytt-\cot\vartheta\Yt+\frac{m^2}{\sin^2\vartheta}\Y\Biggr.\\
\nn
\Biggl. + \frac{2 m}{\sin\vartheta}\left(\Yt-\Y \cot\vartheta\right)
\Biggr].
\eeq
The source \op T_{\ell m}(\omega,r)\cl is given by 
\be
\label{sourceteuk}
T_{\ell m}(\omega,r)=-2r^4~\sqrt{n\left(n+1\right)}~
\tilde{T}^{nn}_{\ell m}(\omega,r)+2\sqrt{n}~
\Delta\Lambda_+~{r^5\over\Delta}
\tilde{T}^{n\bar{m}}_{\ell m}(\omega,r)-{\Delta
\over 2r}\Lambda_+~{r^6\over\Delta}
\Lambda_+r\tilde{T}^{nn}_{\ell m}(\omega,r),
\ee
where
\be
\Lambda_+=\frac{d}{dr_
*}+i\omega,
\ee
and the factors ~$\tilde{T}_{ab}^{\ell m\omega}(r)$~
are the radial parts of Newman-Penrose components of
the stress-energy tensor (see Paper II, app. A.2).
In the present paper we prefer to use another equation, related to the 
BPT equation, which  was derived
by Nakamura and Sasaki \cite{nakamurasasaki}. They  showed that, given
a solution of the inhomogeneous   \bpte, 
there exists a function \op\zns\cl related to \op\ps\cl by
the Chandrasekhar transformation \cite{chandratrasf}
\be
\label{chandratransformation}
\ps = D_{CH} \zns,
\ee
where
\be
\label{chandraoperator}
D_{CH}= \Delta \Lambda_+ \frac{r^2}{\Delta}\Lambda_+ r,
\ee
which satisfies the generalized inhomogeneous \rwe
\be
\label{reggegen}
\Bigg\{
\frac{d^2}{{dr_*}^2}-\Biggl[\omega^2-
\frac{\Delta}{r^5}\Bigl[ \ell(\ell+1)-6M\Bigr]\Biggr]
\Biggr\}
Z_{\ell m}^{NS} =S_{\ell m}^{NS}.
\ee
The source term is related to the source of the BPT equation by
\be
\label{sns}
D_{CH}\left( \frac{r^4}{\Delta}S_{\ell m}^{NS}\right)= -T_{\ell m}.
\ee
We shall now write the source term of the Nakamura-Sasaki equation
explicitely.
If we use, as in paper I, the convenction \op G_{\mu\nu}=2
T_{\mu\nu},\cl
the stress-energy tensor of a point-like mass  \op m_0,\cl
moving along a geodesic \op \underline{r}(\tau)\cl  of the unperturbed
spacetime, is 
\be
T^{\mu\nu}=2m_0 \int{d\tau ~{dz^{\mu}\over d\tau}{dz^{\nu}\over d\tau}
\delta^{(4)}\Bigl(\underline{r}-\underline{r}(\tau)\Bigr)}.
\ee
Since
\[
\delta\Bigl(\bar{r}-r(\tau)\Bigr)=
\sum_i\frac{1}{
\Bigl|\frac{dr}{d\tau}\Bigr|_{\tau_i(\bar{r})}
}
\delta\Bigl(\tau-\tau_i(\bar{r})\Bigr),
\]
where $\tau_{i}(r)$ are the solution of 
\op
r(\tau)=\bar{r},
\cl
the solution can be divided in two branches, 
corresponding to the incoming and outgoing 
part of the trajectory. 
Consequently, the stress-energy tensor can be written as
\be
\label{deltas}
T^{\mu\nu}=2m_0~\sum_{i=1}^{1}{
\frac{2}{r^2\vert \gamma\vert }~
\frac{dx^{\mu}}{d\tau}~
\frac{dx^{\nu}}{d\tau}~
\delta\Bigl(t-t_i(r)\Bigr)
\delta^{(2)}\Bigl(\Omega-\Omega_i(r)\Bigr)},
\ee
where $t_i(r),~\Omega_i(r)$ are the  time
and angular position of the particle on
the i$-$eth branch of the trajectory.
In terms of the tensor (\ref{deltas}), the source of the BPT equation 
can be written as
\beq
&&
T_{\ell m}(\omega,r)
=-m_0~\Biggl\{
\frac{\Delta^2}{r^2} \sqrt{n(n+1)}~
{V'}^2(r)~_0f_{\ell m}(r) e^{i\omega V(r)}+\Biggr.\nn\\
&+&\Biggl.\sqrt{2n}\Delta\frac{\partial}{\partial r_*}
\Bigl(r^2V'(r)~_{-1}f_{\ell m}(r) e^{i\omega V (r)}\Bigr)+
{\Delta\over 2r}
\frac{\partial}{\partial r_*}
\Bigl[\frac{r^6}{\Delta}
\frac{\partial}{\partial r_*}
\Bigl(r~_{-2}f_{\ell m}(r) 
e^{i\omega V(r)}\Bigr)\Bigr]\Biggr\}
e^{-i\omega r_*}.\nn
\eeq
where
\beq
\label{due}
&&V(r)=t(r)+r_*,\\\nn
&&_0f_{lm}(r)=\vert \gamma\vert ~ _0S^*_{\ell m}(\Omega(r)),\\\nn
&&_{-1}f_{lm}(r)=\sigma~
\Biggl(\frac{d\vartheta}{d\tau}-
i\sin{\vartheta}\frac{d\varphi}{d\tau} \Biggr)_{\Omega(r)}
~_{-1}S^*_{\ell m}\Bigl(\Omega(r)\Bigr),\\\nn
&&_{-2}f_{lm}(r)={1\over\vert \gamma\vert}\Bigl(
\frac{d\vartheta}{d\tau}-
i\sin{\vartheta}\frac{d \varphi}{d\tau}\Bigr)^2_{\Omega(r)}
~_{-2}S^*_{\ell m}\Bigl(\Omega(r)\Bigr) ,
\eeq
the prime indicates differentiation with respect to r,  and
\op _{-0}S^*_{\ell m}\Bigl(\Omega(r)\Bigr), _{-1}S^*_{\ell
m}\Bigl(\Omega(r)\Bigr)\cl are the complex conjugate of
the spin-weighted sperical harmonics of weight 0 and 1,
and \op\sigma=sign\Biggl[\frac{dr}{d\tau}\Biggr].\cl
We shall assume that a particle \op m_0\cl 
starts its journey at radial infinity
with energy \op E\cl and angular momentum \op L_z$,
and we shall put $\varphi(r_t)=\tau(r_t)=0$,
at the turning point  \op r=r_t.\cl 
We choose \op E\cl and \op L_z$,
such that the particle is scattered by the star, and follows a geodesic
on the plane \op \vartheta=\frac{\pi}{2},\cl described by the equations
\be
\label{geod}
\frac{dt}{d\tau}=\frac{E}{1-\frac{2M}{r}},~~~~
\frac{dr}{d\tau}\equiv \gamma=\pm\sqrt{ E^2-\left(1-\frac{2M}{r}  \right)
\left( 1+\frac{L_z^2}{r^2} \right)},
~~~~\frac{d\varphi}{d\tau}=\frac{L_z}{ r^2}.
\ee
The source term of the Nakamura-Sasaki equation 
can be derived  from eqs. (\ref{sns}),(\ref{due}) and (\ref{geod}),
as in \cite{nakaooharakoji}.
The result is
\beq
\label{sourceab}
&&S^{NS}_{\ell m}(\omega, r)=\frac{\Delta}{r^5} e ^{-i\omega r_*}
\gbl
\left[ A^1_{\ell m}(\omega, r)+i B^1_{\ell m}(\omega, r)\right]
\\\nn\Biggr.
&&+\Biggl.
\int_r^{\infty}dr'~
\left[ A^2_{\ell m}(\omega, r)+i B^2_{\ell m}(\omega, r)\right]
+\int_r^{\infty}dr'~
\int_{r'}^{\infty}dr''
~\left[ A^3_{\ell m}(\omega, r)+i B^3_{\ell m}(\omega, r)\right]
\gbr
\eeq
where
\beq
A^j_{\ell m}&=\ll a^j_{\ell m}\cos{\omega r_*}
-b^j_{\ell m}\sin{\omega r_*}\rr, \nn\\
B^j_{\ell m}&=\ll b^j_{\ell m}\cos{\omega r_*}
+a^j_{\ell m}\sin{\omega r_*}\rr ,
\label{abgrandi}
\eeq
with j=1,2,3,
\be
\label{abpiccoli}
\cases{
a^1_{\ell m}=\ll K^0_{\ell m}\gamma-
\frac{K^2_{\ell m}}{\gamma}\rr \cos(\omega t-m\varphi),\cr
b^1_{\ell m}=-\bar{C}^1_{\ell m}\sin(\omega t-m\varphi),\cr
a^2_{\ell m}=\ll K^0_{\ell m}\gamma'-
{2\over\gamma r}K^2_{\ell m}\rr\cos(\omega t-m\varphi)+
K^0_{\ell m} {mL_z\over r^2}\sin(\omega t-m\varphi),\cr
b^2_{\ell m}={mL_z\over\gamma r^2}\bar{C}^1_{\ell m}\cos(\omega t-m\varphi),\cr
a^3_{\ell m}=\ll K^0_{\ell m}\frac{\omega mL_z}{\gamma\Delta}
-\frac{2K^2_{\ell m}}{\gamma r^2}\rr\cos(\omega t-m\varphi)+
\frac{2K^0_{\ell m}\omega Mr^2}{\Delta^2}
\sin(\omega t-m\varphi),\cr
b^3_{\ell m}=K^0_{\ell m}\omega
\ll -\frac{2Mr^2\gamma}{\Delta^2}+
\frac{\gamma'r^2}{\Delta}\rr \cos(\omega t-m\varphi)+
\frac{K^0_{\ell m}\omega mL_z}{\Delta}
\sin(\omega t-m \varphi).\cr}
\ee
The constants   \op K_0,...\cl are
\be
\label{cinque}
\cases{
K^0_{\ell m}=-\mu\frac{\sqrt{\lambda (\lambda+2 )}}{\omega^2}~
_0S_{\ell m}^*\left(\frac{\pi}{2} ,0\right)\cr
K^1_{\ell m}=\mu\frac{2iL_z\sqrt{\lambda}}{\omega}~
_{-1}S_{\ell m}^*\left(\frac{\pi}{2},0\right),
\qquad  \bar{C}^1_{\ell m}=-i K^1_{\ell m}\cr
K^2_{\ell m}=
\mu L_z^2~_{-2}S_{\ell m}^*\left(\frac{\pi}{2},0\right),\cr}
\ee
and
\be
\lambda=\ell(\ell+1).
\ee

\section{The solution of the Nakamura-Sasaki equation
and the matching conditions.}
In order to integrate the Nakamura-Sasaki equation (\ref{reggegen})
outside the star, we need to know the value of
\op Z^{NS}_{\ell m}\cl and its derivative at the surface \op r=R.\cl
The Nakamura-Sasaki function is related to the solution
of the Regge-Wheeler equation for the axial perturbations, 
\op Z^{ax},\cl and to that of the Zerilli equation for the polar ones,
\op Z^{pol},\cl by the following relations \cite{ferrarigualtieri}
\be
\label{zns1}
Z^{NS}_{\ell m}(\omega,r)=Z^{1}_{\ell m}(\omega,r)+i Z^{2}_{\ell m}(\omega,r),
\ee
where
\beq
\label{defz1z2}
&&Z^1_{\ell m}(\omega,r) =-\frac{\sqrt{n\left(n+1
\right)}}{4(k-2i\omega\beta)}
\left[(k+2\beta^2 f)Z^{pol}_{\ell m}(\omega,r)
-2\beta Z^{pol}_{\ell m,r_*}(\omega,r)\right],\\
\nn
&&Z^2_{\ell m}(\omega,r) \equiv {\sqrt{n\left(n+1\right)}\over
4i\omega}Z^{ax}_{\ell m}(\omega,r),
\eeq
and 
\be
\label{ztilde}
\cases{
\beta=6M,\cr
k=4n\left(n+1\right),\cr
f=\frac{\Delta}{r^3[(\ell-1)(\ell+2)r+6M]}. &\cr}
\ee
It should be stressed that {\it both} \op Z^1_{\ell m}(\omega,r)\cl and
\op Z^2_{\ell m}(\omega,r) \cl satisfy the Regge-Wheeler equation.
Thus,  \op Z^{NS}_{\ell m}(\omega,R)\cl and its derivative 
can be  found in terms of  \op Z^{pol}(\omega,R)\cl and 
\op Z^{ax}(\omega,R)\cl and their first derivatives.
These values can be found by numerically 
integrating the equations for the polar and axial
perturbations in the interior of the star (cfr.  eqs. \ref{zerfun} and
\ref{axeq}), by imposing the boundary conditions that, 
for each assigned value of the frequency,
the solution is regular  near the origin, and that the perturbation
of the thermodynamical variables vanish at the boundary, so that the
perturbed spacetime reduces to vacuum (cfr. Paper I, sec. 6,7 and 11).
However, these conditions do not define the amplitude of the functions 
\op Z^{pol}(\omega,R)\cl and \op Z^{ax}(\omega,R),\cl
which  depends on the exciting source.
If we name  \op \bar Z^{pol}_{\ell m}(\omega,R)\cl and
\op \bar Z^{ax}_{\ell m}(\omega,R)\cl  the values found by 
integrating  the interior equations with an arbitrary
amplitude at the centre of the star, the `true' values of
\op Z^{pol}(\omega,R)\cl and \op Z^{ax}(\omega,R)\cl  will be
\be
\label{conda}
\cases{
Z^{pol}_{\ell m}(\omega,R)=\chi^1_{\ell m}(\omega)~
\bar Z^{pol}_{\ell m}(\omega,R)\cr
Z^{ax}_{\ell m}(\omega,R)=\chi^2_{\ell m}(\omega)~
\bar Z^{ax}_{\ell m}(\omega,R)\cr},\qquad
\ee
and \op \chi^1_{\ell m}(\omega)\cl and \op \chi^2_{\ell m}(\omega)\cl
have to be found.
Consequently, the functions \op Z^1_{\ell m}(\omega,R)\cl and 
\op Z^2_{\ell m}(\omega,R)\cl  defined in eqs. (\ref{defz1z2}) suffer 
the same ambiguity
\be
\label{cond1}
\cases{
Z^{1}_{\ell m}(\omega,R)=\chi^1_{\ell m}(\omega)~\bar Z^1_{\ell m} (\omega,R)\cr
Z^{2}_{\ell m}(\omega,R)=\chi^2_{\ell m}(\omega)~
\bar Z^2_{\ell m} (\omega,R)\cr},\qquad
\ee
and $\bar Z^1_{\ell m}(\omega,R),\bar Z^2_{\ell m} (\omega,R)\cl are 
(cfr. eqs. \ref{defz1z2})
\beq
\label{defz1bar}
&&\bar Z^1_{\ell m} (\omega,R)=-\frac{\sqrt{n\left(n+1
\right)}}{4(k-2i\omega\beta)}
\Biggl[(k+2\beta^2 f)\bar Z_{\ell m}^{pol}\Bigl(\omega,R\Bigr) -
2\beta\bar  Z^{pol}_{\ell m,r_*}\Bigl(\omega,R\Bigr) \Biggr],\\
\label{defz2bar}
&&\bar Z^2_{\ell m}(\omega,R)= {\sqrt{n\left(n+1\right)}\over
4i\omega}\bar Z_{\ell m} ^{ax}\Bigl(\omega,R\Bigr) ,
\eeq
$\chi^1_{\ell m}(\omega)\cl and 
\op \chi^2_{\ell m}(\omega)\cl
determine how much of the polar and of 
the axial modes  are excited by the particular source 
we are considering.

Eq. (\ref{reggegen})
has to be solved  by imposing that, at the surface,
the solution of the perturbed equations in the interior matches  
continuously with  the exterior
solution, which, in addition, has to
behave as a pure outgoing wave at radial infinity, i.e.
\be
\label{set2}
\cases{
{ L_{RW}} Z^{NS}_{\ell m}(\omega,r) =S^{NS}_{\ell m}(\omega,r)\cr
Z^{NS}_{\ell m}(\omega,R)=Z^1_{\ell m}(\omega,R)+
i  Z^2_{\ell m}(\omega,R)\cr
{Z^{NS}}_{\ell m}^{\prime} (\omega,R)=
{Z^{1}}_{\ell m}^\prime (\omega,R)+i  {Z^{2}}_{\ell m}^\prime
(\omega,R)\cr
Z^{NS}_{\ell m} \sim e^{i\omega r_*}\qquad\qquad r_*
\rightarrow\infty \cr},
\ee
where the prime indicates differentiation with respect to \op r_*,\cl 
and \op{L_{RW}}\cl is the Regge-Wheeler operator
\be
\label{L}
{ L_{RW}}\equiv
\frac{d^2}{dr_*^2} -\Biggl\{\omega^2- 
\frac{\Delta}{r^5}\Bigl[ \ell(\ell+1)-6M\Bigr]\Biggr\}.
\ee
The solution of eqs.
(\ref{set2}) can be found by defining the Green's function appropriate to
the problem, as explained in Paper II. Here we briefly summarize 
the procedure.
For \op r\geq R\cl we define the function
\op Y_{\ell}(\omega,r_*), \op
which satisfies the homogeneous Regge-Wheeler equation 
with the pure outgoing wave condition at infinity
\be
\label{defyz}
\cases{
{L_{RW}}
~ Y_\ell (\omega,r_*)=0&\cr
Y_\ell (\omega,\infty)=e^{i\omega r_*}&\cr},
\ee
and the functions 
\op\bar{Z}^1_{\ell }(\omega,r_*)\cl and \op \bar{Z}^2_{\ell }(\omega,r_*)\cl
that, as \op Y_{\ell}(\omega,r_*), \op 
satisfy a homogeneous Regge-Wheeler equation,
with boundary condition at the surface of the star
given by eqs. (\ref{defz1bar}) and ({\ref{defz2bar}), respectively.
It should be noted that, since the Regge-Wheeler operator does not depend
on \op m\cl, the functions 
\op Y_{\ell}(\omega,r_*),~ \bar Z^1 _{\ell}(\omega,r_*)\cl and 
\op \bar Z^2_{\ell}(\omega,r_*)\cl will be independent of \op m\cl as well.
We then  construct the solution of the Nakamura-Sasaki equation as follows
\beq
\label{solgen}
Z^{NS}_{\ell m}(\omega,r_*)&=&
\frac{\alpha_{\ell m}(\omega)}{W^1_\ell}
\Biggl[
\bar{Z}^1_{\ell}(\omega,r_*)
\int_{r_*}^{\infty}
{Y_\ell(\omega,y_*)S^{NS}_{\ell m}(\omega,y_*)dy_*}\Biggr.\\
\nn
&+&\Biggl. Y_{\ell}(\omega,r_*)
\int_R^{\infty}{\bar{Z}^1_{\ell}(\omega,y_*)
S^{NS}_{\ell m}(\omega,y_*)dy_*}\Biggr]
\\\nn
&+&\frac{\Bigl(1-\alpha_{\ell m}(\omega)\Bigr)}{ W^2_\ell}
\Biggl[
\bar{Z}^2_{\ell}(\omega,r_*)
\int_{r_*}^{\infty}
{Y_\ell(\omega,y_*)S^{NS}_{\ell m}(\omega,y_*)dy_*}\Biggr.\\\nn
&+&\Biggl. Y_\ell(\omega,r_*)
\int_R^{\infty}{\bar{Z}^2_{\ell}(\omega,y_*)
S^{NS}_{\ell m}(\omega,y_*)dy_*}
\Biggr],\label{NS}\nn\\
&&
\eeq
where \op W^1_{\ell }(\omega)\cl and \op W^2_{\ell }(\omega)\cl are 
the wronskians
\beq
W^1_{\ell }(\omega)&=& Y_{\ell }^\prime(\omega,r_*) ~
\bar Z^1_{\ell }(\omega,r_*)-
Y_{\ell }(\omega,r_*) ~ \bar {Z^1_{\ell }}^\prime(\omega,r_*), \\\nn
W^2_{\ell }(\omega)&=& Y_{\ell }^\prime(\omega,r_*) ~
\bar Z^2_{\ell }(\omega,r_*)-
Y_{\ell }(\omega,r_*) ~ \bar {Z^2_{\ell }}^\prime(\omega,r_*),
\eeq
and \op\alpha_{\ell m}(\omega)\cl are constants.
It should be mentioned that, since the source \op 
S^{NS}_{\ell m}(\omega,r_*)\cl given in eqs. 
(\ref{sourceab})-(\ref{cinque}) diverges as \op\gamma^{-1}\cl at the
turning point, the actual integration of
eq. (\ref{solgen}) can be performed by switching to the
proper time \op\tau(r).\cl

It is easy to check that the solution (\ref{solgen}) satisfies the pure
outgoing wave condition at infinity, and that
 the matching condition at the boundary are fulfilled
provided the constants \op \chi^1_{\ell m}(\omega)\cl and 
\op \chi^2_{\ell m}(\omega)\cl are defined as
\beq
\label{defchi}
&&\chi^1_{\ell m}(\omega)=
{\alpha_{\ell m}(\omega) \over W^1_{\ell }(\omega)}
\int_R^{\infty}{Y_{\ell}(y_*,\omega)
S^{NS}_{\ell m}(y_*,\omega)dy_*}
\\
\nn
&&\chi^2_{\ell m}(\omega)=
{[1- \alpha_{\ell m}(\omega)] \over W^2_{\ell }(\omega)}
\int_R^{\infty}{Y_{\ell}(y _*,\omega)
S^{NS}_{\ell m}(y_*,\omega)dy_*}.
\eeq
The solution at radial infinity, to which we are primarily interested,  therefore is
\beq
\label{znsinfty}
Z^{NS}_{\ell m}\ll\infty,\omega\rr=
e^{i\omega r_*}\ll{\alpha_{\ell m}(\omega)
\over W^1_{\ell}(\omega)}\int_R^{\infty}{
\bar{Z}^1_{\ell }(y_*,\omega)
S^{NS}_{\ell m}(y_*,\omega)dy_*}+
\rr.
\\\nn
\ll
{\ll 1-\alpha_{\ell m}(\omega)\rr\over W^2_{\ell}(\omega)}\int_R^{
\infty}{\bar{Z}^2_{\ell }(y_*,\omega)
S^{NS}_{\ell m}(y_*,\omega) dy_*}\rr
\eeq
The constants \op \alpha_{\ell m}(\omega)\cl will be determined by the
following  arguments.
Due to the properties of the spherical harmonics 
\beq
&&Y_{\ell m}(\vartheta,\varphi)=
     (-1)^{m}Y_{\ell m}(\vartheta,\varphi+\pi)=
     (-1)^{\ell}Y_{\ell m}(\pi-\vartheta,\varphi+\pi)= 
     (-1)^{\ell+m}Y_{\ell m}(\pi-\vartheta,\varphi),\\\nn
&&Y_{\ell m}(\vartheta,\varphi)=
     (-1)^{m}Y^\ast_{\ell -m}(\vartheta,\varphi),
\eeq
by an inspection of the behaviour of the 
source \op S^{NS}_{\ell m},\cl
it is easy to verify that  the Nakamura-Sasaki function 
satisfies the following identity
\be
\label{simz}
Z^{NS~\ast}_{\ell -m}\left( -\omega,r\right)=
(-1)^\ell Z^{NS}_{\ell m}\left(\omega,r\right).
\ee
and consequently
\beq
\label{simz1z2a}
&&Z^{1\ast}_{\ell -m}(-\omega,R)=(-1)^\ell Z^1_{\ell m}(\omega,R)\\
\nn
&&Z^{2\ast}_{\ell -m}(-\omega,R)=-(-1)^\ell Z^2_{\ell m}(\omega,R).
\eeq
In addition, by explicitely evaluating \op
\delta\Psi_4(t,r,\vartheta,\varphi)\cl for the perturbed metric 
(\ref{pertmetric}), and by using eq.
(\ref{chandratransformation}), we find 
\beq
\label{aaa}
D_{CH} Z^1_{\ell m}&=&
\frac{r \sqrt{n(n+1)}}{4}\Biggl\{
r e^{2\nu}\Bigl[\NR -\LR\Bigr]\Biggr.\\
\nn
&+&\Biggl. r^3 e^{2(\nu-\mu_2)}\left[-\VRPP+
( \mu_2'+\nu'-\frac{2}{r})\VRP\right]\Biggr.\\
\nn
&-&\Biggl. 2 i\omega r^3 e^{(\nu-\mu_2)} \VRP
+r^3\left[\omega^2+
2 i\omega e^{(\nu-\mu_2)} \left(\nu'-\frac{1}{r}\right) \right]\VR 
\Biggr\}
\\\nn
D_{CH} Z^2_{\ell m}&=&
\frac{r \sqrt{n(n+1)}}{4}\Biggl\{
\frac{r}{2}\Biggl[
 e^{2(\nu-\mu_2)}\ha1rp-
e^{2(\nu-\mu_2)}(\mu_2'+\nu')\h1r+
i \omega e^{(\nu-\mu_2)}
\h1r\Biggr.\Biggr.\\
\nn
&-&\Biggl.\Biggl.
e^{(\nu-\mu_2)}\hhrp+2\nu'e^{(\nu-\mu_2)}\hhr
-i \omega\hhr \Biggr]
\Biggr\}
\eeq
From the property of the metric perturbations given in
eq. (\ref{prop2}), and since
\op
D^*_{CH}(\omega)=D_{CH}(-\omega)
\cl
it follows that
\beq
\label{fff}
&& Z_{\ell m}^{1 *}(\omega,r)=(-1)^{m} Z_{\ell -m}^{1}(-\omega,r)\\
\nn
&& Z_{\ell m}^{2 *}(\omega,r)=(-1)^{m} Z_{\ell -m}^{2}(-\omega,r).
\eeq
Eqs. ( \ref{simz1z2a}) and (\ref{fff}) are compatible 
only if
\beq
\label{compat}
&&(-1)^\ell Z^1_{\ell m}(\omega,R)=(-1)^m Z^1_{\ell m}(\omega,R)\\\nn
&&-(-1)^\ell Z^2_{\ell m}(\omega,R)=(-1)^m Z^2_{\ell m}(\omega,R),
\eeq
which imply that
\begin{itemize}
\item{}
if \op (\ell +m)\cl is even, ~~~\op Z^2_{\ell m}(\omega,R)=0,\cl
\item{}
if \op (\ell +m)\cl is odd, ~~~\op Z^1_{\ell m}(\omega,R)=0,\cl
\end{itemize}
From eqs. (\ref{cond1}) we see that if \op (\ell +m)\cl is even, in order
\op Z^2_{\ell m}(\omega,R)\cl to be zero \op \chi^2_{\ell m}(\omega)\cl
must vanish, which means that \op \alpha_{\ell m}(\omega)=1\cl (see eq.
\ref{defchi}). Similarly, if \op (\ell +m)\cl is odd \op \chi^1_{\ell
m}(\omega)\cl must vanish, i.e. \op \alpha_{\ell m}(\omega)=0.\cl 
Thus,  the complete solution of the Nakamura-Sasaki equation at
radial infinity is (see eq. \ref{znsinfty})
\begin{itemize}
\item{}  \op (\ell +m)\cl even
\be
\label{finalsola}
Z^{NS}_{\ell m}\ll\omega,\infty\rr=
{e^{i\omega r_*}
 \over W^1_{\ell }(\omega)}
\int_R^{\infty}{
\bar{Z}^1_{\ell }(\omega,y_*)
S^{NS}_{\ell m}(\omega,y_*)dy_*},
\ee
\item{}  \op (\ell +m)\cl  odd
\be
\label{finalsolb}
Z^{NS}_{\ell m}\ll\omega,\infty\rr=
{e^{i\omega r_*}
\over W^2_{\ell }(\omega)}\int_R^{\infty}{
\bar{Z}^2_{\ell }(\omega,y_*)
S^{NS}_{\ell m}(\omega,y_*)dy_*}.
\ee
\end{itemize}
Equations (\ref{finalsola}) and (\ref{finalsolb}) show that
the polar and the axial perturbations contribute to the solution \op
Z^{NS}_{\ell m}\cl depending on \op (\ell +m)\cl being even or odd,
respectively. This fact is not surprising.  For example,
if we consider  the non-homogeneous Regge-Wheeler equation
written for the  process under consideration
\be
{\bf L_{RW}} Z^{RW}_{\ell m}(\omega,r) =S^{RW}_{\ell m}(\omega,r),
\ee
where the source explicitely is
\begin{eqnarray}
S^{RW}_{\ell m}&=&\frac{4\imath m_0 e^{\nu-\mu_2}}{\sqrt{2}}
             e^{\imath \omega[t(r)]} \left\{\imath e^{2\nu}
             \frac{\sqrt{2} n(\ell)}{r^2(r-2M)}\,{L_z}
             \ Y^{*}_{\ell m,\vartheta}\mid_{\vartheta=\pi/2}
             + \right. \nonumber \\
          && - 2  e^{2\nu} m \frac{\sqrt{2} m(\ell)}{r^{4}}
             \,g(r)^{-1/2}\,{L_z}^{2} \
             Y^{*}_{\ell m,\vartheta}\mid_{\vartheta=\pi/2}+ \nonumber \\
          && -\frac{m(\ell)}{r} m {\tilde L}^{2} \
             Y^{*}_{\ell m,\vartheta}\mid_{\vartheta=\pi/2} \left[
             \frac{d}{dr}\left(r^2e^{2\nu}\frac{\sqrt{2} }{r^{4}}
             \,g(r)^{-1/2}\right)+ \right. \nonumber \\
          && \left. \left. +r^2e^{2\nu}\frac{\sqrt{2} }{r^{4}}
             \,g(r)^{-1/2}\Bigl(\imath \omega \frac{dt(r)}{dr}- \imath m
             \frac{d\varphi(t(r))}{dr}\Bigr) \right] \right\},
\end{eqnarray}
and
\op
g(r)=\displaystyle{\left(\frac{2M}{r}-
          \frac{r-2M}{r^3}{L_z}^{2}\right)}, \cl
\op n(\ell)=[\ell(\ell+1)]^{-1},\cl and 
\op m(\ell)=[\ell(\ell+1)(\ell+2)(\ell-1)]^{-1}, \cl
when  \op (\ell +m)\cl is even  
\op
\ Y^{*}_{\ell m,\vartheta}\mid_{\vartheta=\pi/2}=0,\op
the source term is zero,  and the axial perturbations are not excited. 

It should be mentioned that the procedure used in this paper
to find the constants \op \alpha_{\ell m}(\omega)\cl exploits 
the symmetry properties of the source of the Nakamura-Sasaki
equation, and is much simpler than that described in Paper II.

\section{The model of star}
We have integrated eqs. (\ref{poleq}) and (\ref{axeq})
in the interior of a star with a polytropic equation of state
\op p=K\rho^{n},\cl with \op K=100~km\cl and
\op n=2.\cl  If the  central density  is chosen to be
\op \rho_c=3\cdot 10^{15}~g/cm^3,\cl
the radius and mass of the star are, respectively, \op R=8.86~km,\cl and
\op M=1.266~M_\odot,\cl with a ratio \op R/M=4.7\cl
For this model of star, in Fig. 1 we plot the resonance curve evaluated
for the \op\ell=2,\cl polar perturbations, which allows to locate the
frequencies of the quasi-normal modes of the star.
This curve is obtained by integrating the polar equations inside and
outside the star (eqs. \ref{poleq},\ref{zerfun}) 
for assigned real values of the 
frequency, and by evaluating the amplitude of the standing wave,
\op\left[\alpha^2(\omega)+\beta^2(\omega) \right],\cl  
prevailing at radial infinity.
It can be shown that the values of frequency at which
this curve  exhibits a sharp minimum correspond to the real
part of the complex eigenfrequency of a quasi-normal mode, 
provided the corresponding imaginary part 
is small enough ($\omega_i << \omega_0.$) 
\cite{chandraferrari1},\cite{roland} .
The damping time associated to a mode is related
to the curvature of the parabola that fits the curve
near a minimum:  smoother minima correspond
to  shorter damping times.

As in newtonian theory, the classification of the polar modes
is based on the behaviour of the perturbed fluid according to the
restoring force that is prevailing \cite{cowling}: the g-modes, or gravity
modes, when the force is due to the eulerian change in the density,
the p-modes, when it is due to a change in pressure, and
the f-mode, that  is the
generalization of the only possible mode of oscillation of an
incompressible homogeneous sphere \cite{fmode}.
An inspection of the  behaviour of the thermodynamical
variables  in correspondence of the minima of the resonance curve shown in
Fig. 1, allows to identify the corresponding modes.
The resonance curves for \op \ell=3\cl and \op\ell=4\cl 
are plotted in Fig. 2.
A resonance curve can be  computed also for the axial perturbations by
solving eq. (\ref{axeq}). For the model of star we consider, it exhibits a
monotonically decreasing behaviour, showing that no slowly damped axial
modes exist in this case.
The algorithm based on the resonance curve
allows to find only  the slowly damped modes
($\omega_i << \omega_0.$). 
To find the highly-damped modes (the w-modes \cite{kokkotasschutz92})
other methods have to be used.
In table 1, we  tabulate the values of the complex 
frequency of the first few polar
w-modes computed in ref. \cite{kokkotasschutz92}
for the same model of star  considered in the present paper.
We do not go into more details about these modes because, as we shall see,
they are not excited in  scattering processes.

\section{Numerical results}

As discussed in section IV, 
we  find \op \bar Z_\ell^{pol}(\omega,R)\cl and 
\op\bar  Z_\ell^{ax}(\omega,R)\cl  and their first derivatives
by solving the  interior equations (\ref{poleq}) and (\ref{axeq})
for assigned values of the frequency. 
With these values we construct \op\bar Z^1_\ell(\omega,R)\cl
and \op\bar Z^2_\ell(\omega,R),\cl and
integrate  the homogeneous Regge-Wheeler equation for these functions 
and for the function \op Y^1_\ell(\omega,r_*)\cl defined in
eq. (\ref{defyz}), up to radial infinity.
These functions are needed to construct
the solution of the Nakamura-Sasaki equation,  and find
its asymptotic behaviour 
given by eqs. (\ref{finalsola}) and (\ref{finalsolb}).
The perturbations are assumed to be excited by a massive particle \op
m_0,\cl which is scattered by the gravitational field of the star;  we
set  the angular momentum and the energy respectively to
\op L_z=5,\cl  and \op E=1.007.\cl
For these values,  the  turning point is located at \op r_t/M=9.2\cl
(case a).
We then repeat the calculations allowing the particle to get closer
to the star: we choose \op E=1.097,\cl so that \op r_t/M=5.0\cl
(case b).

For a plane wave (see for example \cite{MT} pg. 522 eq. (488))
\be
\label{pshh}
\delta\Psi_4(t,r,\vartheta,\varphi)=
-\frac{1}{2}\ll \ddot{h}_{\varphi\varphi} (t,r,\phi,\theta)+i
\ddot{h}_{\vartheta\varphi} (t,r,\phi,\theta)\rr,
\ee
and since the \op (0r)$-component of the
 stress-energy pseudotensor  can be written as
\be
\label{stress}
t_{0r}= \left( \frac{dE}{dS dt}\right) =
{1\over {16\pi }}\left\{ \left[ \dot h_{\varphi\varphi}
(t,r,\varphi,\vartheta)\right]^2+
\left[ \dot h_{\vartheta\varphi}(t,r,\varphi,\vartheta)\right]^2\right\},
\ee
it follows that the energy emitted in gravitational waves per unit solid
angle and unit frequency is
\be
\label{dedome1}
\frac{dE}{d\Omega d\omega}=\frac{r^2}{2\omega^2}
\vert \delta\Psi_4(\omega,r,\vartheta,\varphi)\vert^2.
\ee
By integrating over the solid angle, and by using
the relation existing between \op \delta\Psi_4\cl and the
Nakamura-Sasaki function and discussed in section III,
it can be shown that the energy emitted per unit frequency 
can be expressed as a function of the amplitude of \op Z^{NS}_{\ell m}\cl
at infinity, i.e.
\be
\frac{dE_{\ell m}}{d\omega}=
16\omega^2\vert A^{NS}_{\ell m}(\omega)\vert^2,\qquad\qquad
0 < \omega < \infty,
\ee
where
\be
Z^{NS}_{\ell m}(\omega,\infty)
~\rightarrow~ A^{NS}_{\ell m}(\omega)e^{i\omega r_*},\ee
and \op A^{NS}_{\ell m}(\omega)\cl is determined by numerical
integration.
We find that, as in the case of the scattering of masses by a black hole,
the energy spectrum of the emitted gravitational radiation
is contributed mainly by the \op\ell=m$ component.

In Fig. 3 we plot the \op\ell=m=2\cl  energy spectrum 
emitted in case a, when the turning point is \op r_t=9.2~M.$
It  is interesting to compare this spectrum with that obtained 
when a black hole scatters a small mass
(see \cite{nakaooharakoji} for an extensive review).
In that case  the energy  is emitted essentially by the scattered mass,
and most of it is radiated when the mass
transits through the turning point. 
Indeed, the energy spectrum is peaked at a
frequency which is related to the angular velocity of the mass 
at the turning point
\be
\omega_{r_t}=\ell\Bigl( \frac{d\varphi}{dt}\Bigr)_{r=r_t}.
\ee
Thus, the black hole quasi-normal modes are not excited in these processes.
In our case a, and for \op\ell=2,\cl the frequency corresponding
to the angular velocity of the mass at the periastron is
\op \omega_{r_t}M=0.092\cl 
and the spectrum shown in Fig. 3
exhibits a sort of peak close to that frequency,
showing that part of the energy is still emitted by the mass as
a synchrotron radiation. However the very distinctive feature of that
spectrum is a  very sharp peak which occurs at the
frequency of the fundamental f-mode (cfr. Fig. 1). If we allow the 
mass to  get closer to the star, as shown in Fig. 4 for case b
when the turning point is \op r_t=5,\cl the amplitude of the 
f-mode peak  increases by more than a factor \op 10^2,\cl 
and new peaks appear, corresponding to the excitation of the p-modes.
(In this case \op \omega_{r_t}M=0.221$).
Thus, we can conclude that, unlike the case of black holes, the polar
quasi-normal modes of the star can be excited in scattering processes,
to an extent that depends on how close the mass can get to
the star. Conversely, from table 1 and Fig. 3 and 4,  
we see that the w-modes are not significantly excited in these processes.
It should be mentioned that,
although  we plot the energy spectra  for \op \omega M \leq 1,\cl 
we have extended our numerical integration at higher frequencies
finding that the signal is negligible compared with that  shown in the
figures and no further modes appear.
In Fig. 5 we plot the energy spectrum for \op \ell=2\cl and  different
values of \op -\ell \leq m \leq \ell.\cl We see  that when \op\ell+m\cl
is odd, no modes are excited, because the spectrum is contributed
only by the axial perturbations which 
are not resonant for this model of star.
In Fig. 6 we compare the energy spectrum emitted for
\op\ell=2, 3\cl and  4.
The excitation of the quasi-normal modes is exhibited by all multipoles.

\section{Concluding remarks}

In 1983 Lindblom and Detweiler investigated 
the relation existing between the frequencies of 
oscillation  of stars
and the equation of state (EOS) prevailing in the interior,
showing that quasi-normal modes carry information
on the internal structure of a star \cite{lindblomdetweiler83}.
They computed the frequency of the fundamental mode of
oscillation  (the f-mode),
for several EOS's, and
more recently this study has been extended to include a few modern
EOS's suggested for neutron stars, for which also
the frequencies of the first p-mode and of the polar and axial w-modes
have been computed.  Besides, it has been
shown that the knowledge of these frequencies allows to
infer empirical relations between the mode
frequency and the macroscopic parameters of the star: the mass and the
radius \cite{anderssonkokkotas98},\cite{bertibenharferrari}.
Thus, there is no doubt that the observation of the quasi-normal modes
would provide relevant information on
both the structure of neutron stars and on the nature of the
nuclear interactions  at supernuclear densities.

The real problem is to ascertain whether these modes can
be excited in realistic astrophysical situations,
and what is the amount of gravitational energy that is
emitted at the corresponding frequencies.
The work presented in this paper is a first step in this direction.
The energy spectra  plotted in Fig. 3-6 clearly show that the
quasi-normal modes of stars can be excited in scattering processes.
The waveforms  of the corresponding signals
are essentially  exponentially damped, pure sinusoids, at the frequency
of the f-mode of the considered star (\op\nu\sim 3~kHz, \tau\sim 0.07~s$),
and with an amplitude which scales as
\op m_0^{-2}$, and depends on how close the scattered mass
gets to the star.

The p-modes can be excited if the mass gets sufficiently close
(Fig. 4), but the f-mode contribution is
likely to be the dominant one.
Thus, the signal emitted by a star excited by a scattered mass 
is  a pure note, emitted at the frequency, and with the damping time, of
the fundamental f-mode.
The spectra we compute do not show the excitation of the w-modes.
However, they may be excited in other processess, like for example
 the capture of masses by stars.

We plan to extend the present work, to compute the energy and the
waveforms emitted by masses orbiting around compact stars.

\acknowledgments

We gratefully acknowledge Dr. Emanuele Leonardi and the whole INFN CONDOR
group for introducing us to CONDOR, a procedure which allowed us to 
access INFN computational facilities distributed throughout Italy.

\newpage

\begin{figure}
\caption{
The \op\ell=2$- polar resonance curve
is plotted versus the normalized frequency 
\op\omega M$.
Each minimum corresponds to a quasi-normal mode, and the modes are
labelled according to the Cowling classification.
}
\label{fig1}
\end{figure}

\begin{figure}
\caption{
The \op\ell=3$ (right) and \op \ell=4$ (left)  polar resonance curves
are plotted as a  function of the frequency as in fig. 1.
}
\label{fig2}
\end{figure}

\begin{figure}
\caption{
The \op\ell=m=2$ - component of the
energy spectrum of the gravitational radiation
emitted  when a mass \op m_0,\cl with
\op L_z=5\cl and \op E=1.007,\cl is scattered by a polytropic star,
is plotted versus the normalized frequency.
In this case the turning point is located at \op r_t/M=9.2$ (case a).
}
\label{fig3}
\end{figure}

\begin{figure}
\caption{
The \op\ell=m=2$ - component of the  energy spectrum 
emitted  when the scattered mass \op m_0\cl 
has an energy at infinity, \op E=1.097,\cl
higher than that considered in fig. 1.
In this case the turning point is located at \op r_t/M=5.0,\cl
and the mass gets closer to the star (case b).
}
\label{fig4}
\end{figure}

\begin{figure}
\caption{
The $m$-components of the  energy spectrum 
are plotted for \op\ell=2\cl for case b.
We see that when \op(\ell+m)\cl is even, the polar quasi-normal modes are
excited. When \op(\ell+m)\cl is odd, the energy is emitted essentially by
the scattered mass (see text).
}
\label{fig5}
\end{figure}

\begin{figure}
\caption{
The energy spectrum emitted at \op\ell=2\cl is compared with that emitted
by higher multipoles (case b).
}
\label{fig6}
\end{figure}
\newpage

\begin{table}
\caption{
The values of the real and imaginary part of the 
frequencies of the first few polar
w-modes  for the model of star under consideration.
\label{table1}}
\begin{tabular}{lcr}
$N$ & $Re(\omega M)$  & $Im(\omega M)$ \\
\tableline
1 &  0.53 &   0.30\\
2 &  0.88 &   0.38\\
3 &  1.22 &   0.43\\
4 &  1.57 &   0.47\\
5 &  1.91 &   0.50\\
\end{tabular}
\end{table}

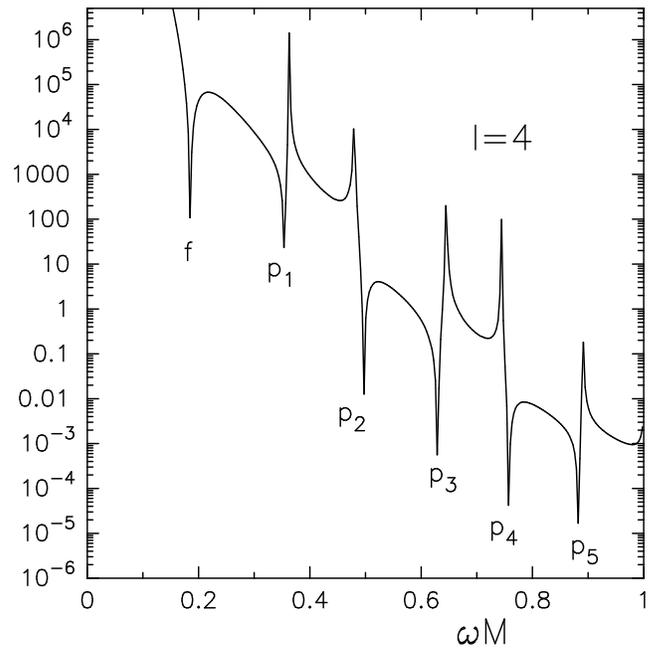
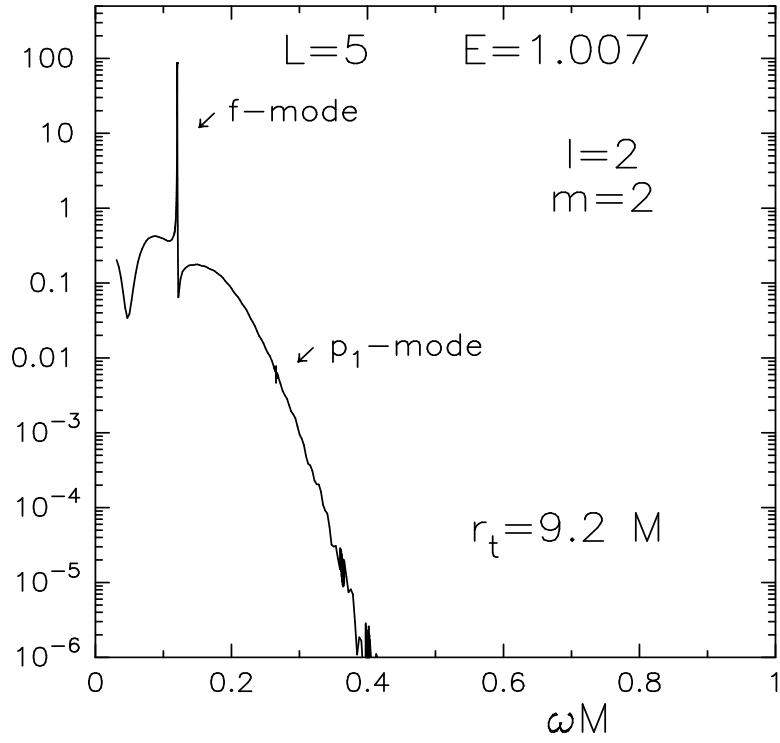
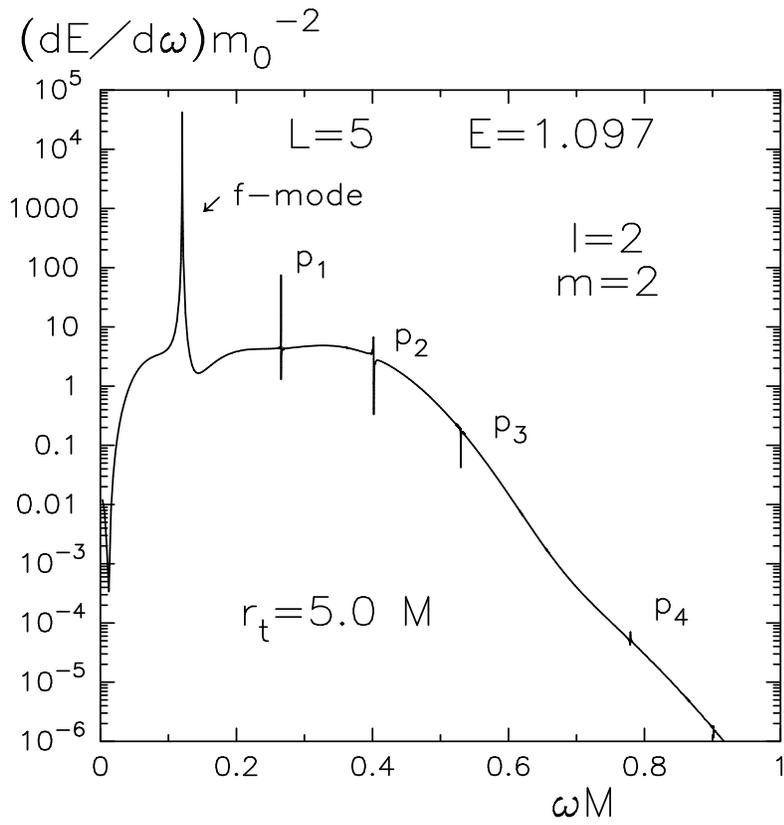
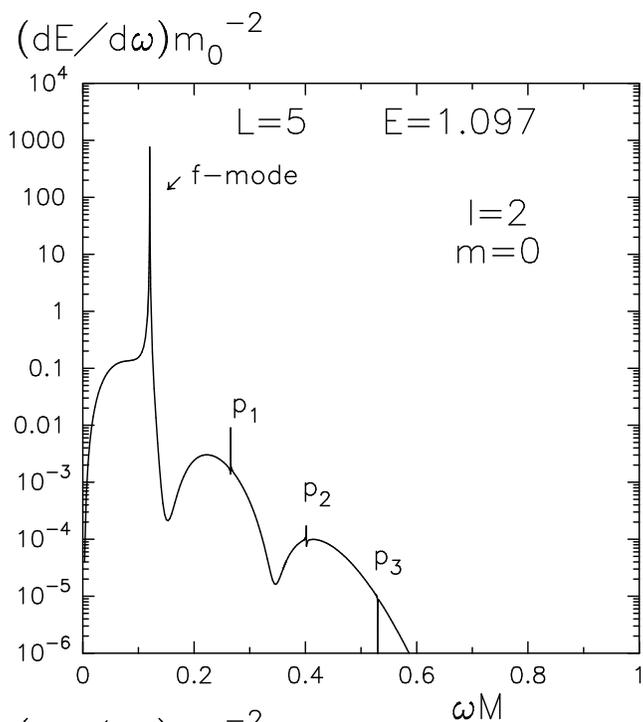
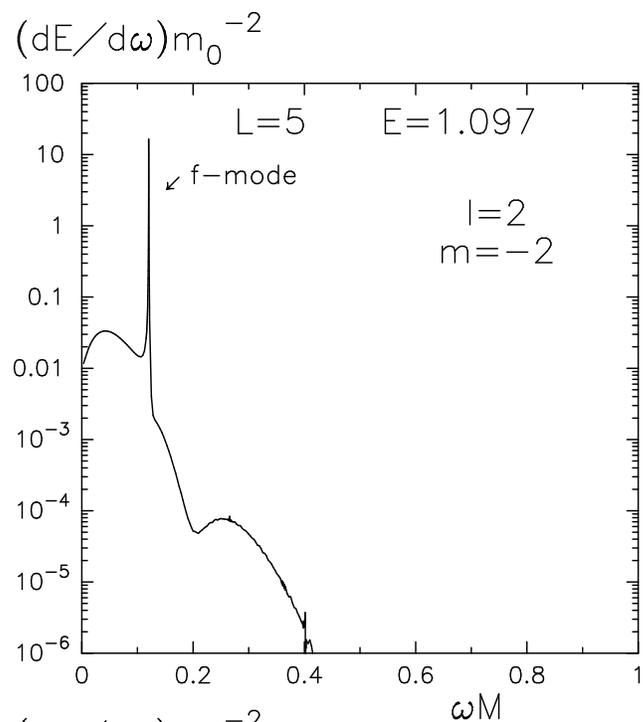
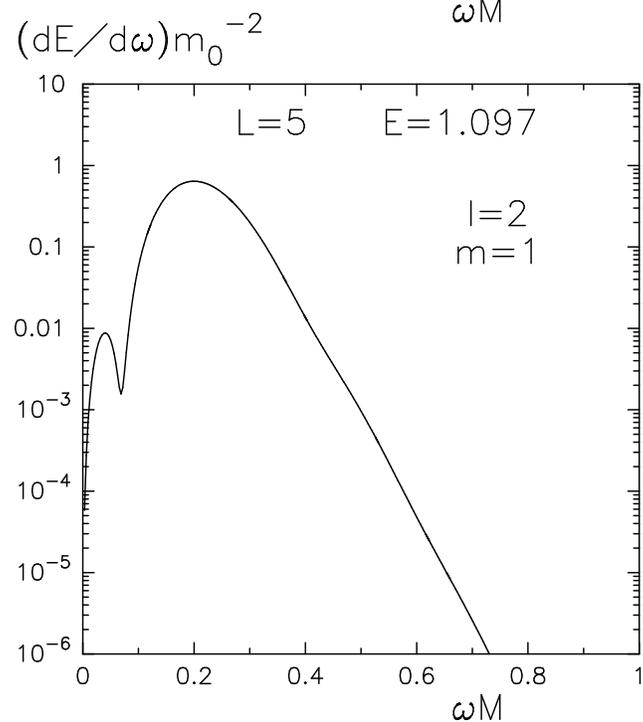
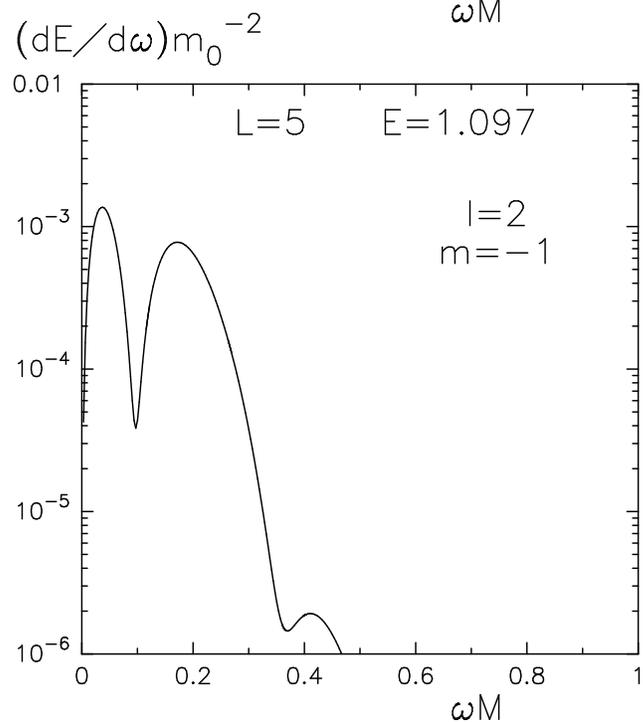
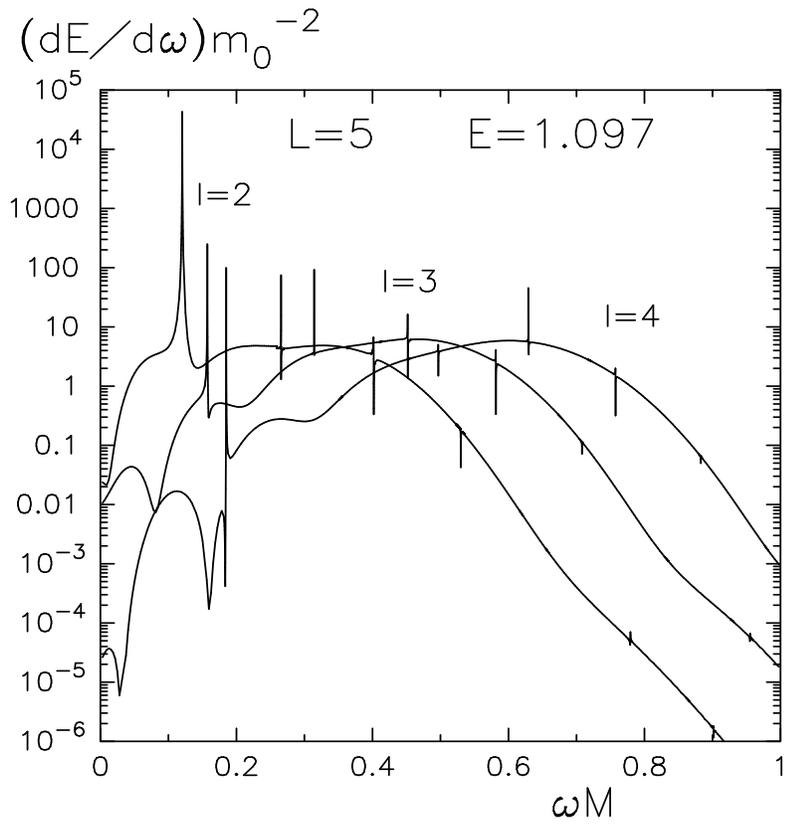
\end{document}